\documentclass[openacc]{rsproca_new}

\titlehead{Research}
\usepackage[sort&compress, numbers]{natbib}
\usepackage{lineno}
\begin{document}
\definecolor{altgreen}{RGB}{46, 139, 87}

    	\title{Evolution of Non-Stationary Quasi-Periodic Pulsations in Solar Flares with Aditya-L1/HEL1OS and GOES}
	\author{
		Srikar Paavan Tadepalli$^{1,2}$,  Abhishekh Kumar Srivastava$^{2}$ and Sankarasubramanian K$^{1}$
	}
	
	\address{$^{1}$ U R Rao Satellite Centre, Bengaluru, India\\
		$^{2}$Department of Physics, IIT-BHU, Varanasi, U.P., India\\
	}
	
	\subject{xxxxx, xxxxx, xxxx}
	
	
\corres{Srikar Paavan Tadepalli\\
	\email{srikar@ursc.gov.in}}

\begin{abstract}

	In the present work, we use a phase-resolved methodology for the detection and characterization of non-stationary QPPs in the hard X-ray lightcurves observed from the HEL1OS instrument onboard Aditya-L1 mission, complemented by cross-examination using soft X-ray observations from GOES. Our approach combines variational mode decomposition with Hilbert–Huang analysis and wavelet transforms to sharpen transient QPP signatures, extract narrowband amplitude–frequency modulated modes, and track instantaneous period drift across various phases of the flares. By applying these tools to preflare, impulsive, and decay  phases and associated intervals, we identify the timing, duration, and evolution of statistically significant oscillations and assess their relation to thermal and non-thermal flare components. Finally, we extend the method to the catalogue of non-stationary QPP flares reported by Mehta et al. (2023), where we identify statistically significant QPPs in both impulsive and decay phases for 74 out of 98 flares. Among these, 56.8\% exhibit small period drifts, while 43.2\% show larger deviations. Positive period drift dominates (73\%), indicating a systematic increase in characteristic timescales as flares evolve. A phase–period–duration analysis reveals distinct regimes of non-stationarity, including overlapping multi-mode oscillations, intermittent same-period episodes, and strongly evolving multi-modal behaviour.
    
\end{abstract}

\maketitle

\section{Introduction}
\label{sect:intro}

Quasi-periodic pulsations (QPPs) are among the most intriguing and ubiquitous signatures in solar and stellar flare emissions, manifesting as quasi-regular modulations in light curves across the electromagnetic spectrum from radio and microwave to soft and hard X-rays and gamma rays \cite{NakariakovMelnikov2009,VanDoorsselaere2016,Zimovets2021}. Reported periods range from fractions of a second to several minutes, with amplitudes ranging from a few percent to order-of-magnitude excursions above the background level \cite{NakariakovMelnikov2009,Kupriyanova2020}. Because QPPs imply repetitive or oscillatory processes in the flare energy release, they have long been regarded as potential diagnostics of the physical mechanisms operating in magnetised plasma environments. Their ubiquity across both solar and stellar flares suggests that they reflect fundamental properties of magnetic energy conversion and particle acceleration \cite{VanDoorsselaere2016,Zimovets2021}.

Early studies of QPPs focused primarily on detecting periodic signatures using Fourier and Lomb–Scargle periodograms. These approaches identify peaks above assumed white-noise backgrounds, but flare light curves are now known to exhibit strong coloured (red) noise behaviour with power-law Fourier spectra \cite{Vaughan2005,Inglis2015}. Such noise characteristics can produce spurious spectral peaks if not properly accounted for, making robust statistical significance testing essential in QPP searches. 
Wavelet analysis rapidly gained prominence because of its ability to localise oscillatory power in both time and frequency. The widely adopted framework developed by Torrence and Compo \cite{TorrenceCompo1998} provides explicit significance testing against red-noise backgrounds, allowing transient periodicities to be identified with greater reliability. However, wavelet spectra can still be affected by strong flare trends and by the cone of influence at long periods, while global integration of wavelet power averages over time and may therefore obscure evolving oscillatory behaviour.

A key limitation of many traditional approaches is the implicit assumption of stationarity: Fourier-based approaches and wavelet analysis, are ill-suited to capture non-stationary QPP signatures because they assume a fixed or globally averaged oscillatory pattern. In practice, many QPPs exhibit pronounced non-stationarity, with oscillatory trains lasting only a few cycles and with evolving period, amplitude, and phase \cite{Nakariakov2018}. Systematic statistical surveys have confirmed that this behaviour is common. Mehta et al. (2023) \cite{Mehta2023} found that a majority of QPPs in a sample of M- and X-class flares show measurable period evolution between flare phases, indicating that non-stationarity is a fundamental property of QPP signals. This prevalence has been supported by complementary statistical studies that emphasise the need to recognise and quantify departures from stationarity in survey-scale analyses \cite{Pugh2017}.

To address these challenges, adaptive decomposition techniques have increasingly been applied to flare time series. Empirical Mode Decomposition (EMD) and its variants allow non-linear and non-stationary signals to be decomposed into intrinsic mode functions without assuming a fixed basis \cite{Huang1998}. These approaches have been used successfully in several solar flare studies \cite{Kolotkov2015,Nakariakov2018}. However, EMD-type methods can suffer from mode mixing, in which physically distinct oscillations contaminate each other. 

An alternative is Variational Mode Decomposition (VMD), an optimisation-based approach that enforces spectral compactness on extracted modes, yielding narrowband, non-overlapping components that are more robust to noise and better suited for subsequent Hilbert-Huang Transform (HHT) spectral analysis. Thus, VMD enables estimation of instantaneous frequencies and amplitudes that are physically interpretable for non-stationary oscillations, addressing some of the limitations of both wavelets and EMD for characterising complex temporal evolution \cite{Dragomiretskiy2014}.

Refinements to traditional methods have also been pursued. Transformations of autocorrelation functions before wavelet analysis can enhance ridge localisation and significance of damped oscillations \cite{Nakariakov2018}, and synchrosqueezed wavelet transforms have been introduced in related astrophysical contexts to sharpen instantaneous frequency estimation \cite{RamnBallesta2022}. Bayesian and Gaussian process modelling approaches have been applied to parameterise non-stationary damped sinusoidal components and provide statistically rigorous model comparisons against red-noise alternatives, though at significant computational cost which currently limits their application to detailed case studies rather than large surveys.

Comparative studies have demonstrated that no single technique is sufficient for robust QPP detection \cite{Broomhall2019}. Instead, multi-method confirmation - combining power spectral analysis, wavelets, autocorrelation methods, and adaptive decompositions - provides the most reliable identification of oscillatory behaviour in flare light curves \cite{Broomhall2019,Szaforz2025}. Despite the variety of techniques, the literature converges on a central conclusion: robust detection of QPPs requires methods that can track non-stationary, multi-modal oscillations and assess their significance against realistic coloured noise backgrounds.

In this study, we apply a time–resolved method for detecting and characterizing non-stationary QPPs in solar flare lightcurves. Section \ref{subsect:data} describes the observational data used in this work, including hard X-ray measurements from the HEL1OS CdTe and CZT detectors and the corresponding soft X-ray lightcurves from GOES. Section \ref{subsect:methods} outlines the analysis framework, which combines variational mode decomposition with Hilbert–Huang ridge tracking and wavelet analysis to isolate transient oscillatory components and follow their instantaneous period evolution. The results are presented in two stages, in Section \ref{sect:analyss_and_results}. First, representative case studies from HEL1OS observations are analysed to illustrate different manifestations of QPP non-stationarity across multiple energy bands. Second, the same methodology is applied to the catalogue of non-stationary QPP flares identified by Mehta et al. (2023) in order to investigate the statistical properties of period evolution across flare phases. The combined analysis reveals distinct regimes of non-stationary behaviour and provides insight into the physical processes responsible for the temporal organisation of flare QPPs. The broader implications of these findings for the interpretation of QPP mechanisms and flare energy release are discussed in Section \ref{sect:disc}.

\section{Data and Methods}
\label{sect:data_and_methods}

\subsection{Data}
\label{subsect:data}

The dataset utilized in this study is from the HEL1OS - High Energy L1-Orbiting X-ray Spectrometer \cite{Nandi2025} instrument onboard Aditya-L1 \cite{Parate2025,Sankarasubramanian2025b}, which was launched by Indian Space Research Organization (ISRO) in September, 2023. HEL1OS is a Sun-as-a-star observing hard X-ray spectrometer that operates in the energy range of 8 - 150 keV, aimed at studying solar flares. The instrument's energy coverage and sensitivity are achieved by housing two sets each of two types of detectors - Cadmium Telluride (CdTe) and Cadmium Zinc Telluride (CZT), which cover the energy ranges 8 - 60 keV and 20 - 150 keV respectively. Further details on the instrument, system configuration, and specifications are available in \cite{Nandi2025}. The instrument provides both an event list and spectral data as part of Level 1 data. The temporal resolution of the event list is $10~\text{ms}$. For the purposes of this work, where the aim is to characterise QPPs across the different phases of a flare, we restrict ourselves to the lightcurves produced by the broad energy channels of the CdTe and CZT detectors, integrated to a cadence suitable for resolving oscillations on timescales of a few seconds to several minutes. Though it is feasible to have sub-second cadences, in this work we use lightcurves from both detectors with a time cadence of 4 seconds. To analyse the oscillations in the thermal and non-thermal segments of the X-ray flux, band-wise lightcurves are segregated into the following: CdTe lightcurves are divided into separate bands  [10 - 15 keV], [15 - 20 keV ], [20 - 25 keV],  and the CZT lightcurves are also divided into bands [25 - 40 keV],[40 - 70 keV],[70 - 120 keV].

To extend the HEL1OS results into the soft X-ray context, X-ray irradiance data from the Geostationary Operational Environmental Satellites (GOES) are used as well. Specifically, we employ the science-ready level-2 flux products from the X-ray Sensor (XRS) at one-second cadence, which have become the community standard for flare timing and classification. These data provide soft X-ray fluxes in the 1–8 \AA \   [1.5–12.4 keV] and 0.5–4 \AA \ [3–24 keV] bands. The one-second cadence lightcurves are sufficient to resolve oscillations on tens-of-second timescales, and are well suited for tracking the flare trend and thermal component evolution. In this study, the GOES data are used in two complementary ways: first, to provide an absolute timing reference for the HEL1OS hard X-ray observations, ensuring that phase boundaries such as flare onset and peak are consistently defined; and second, to act as an independent channel in which to search for quasi-periodic signatures at lower photon energies, where thermal processes dominate. The combined use of HEL1OS and GOES thus gives a multi-channel dataset covering both non-thermal and thermal regimes. HEL1OS provides high time-resolution, photon-counting lightcurves in the hard X-ray range, ideally suited to probing impulsive-phase QPPs and their non-stationary evolution, while GOES offers simultaneous soft X-ray diagnostics of the flare's thermal component. By analysing both, we ensure that the QPP detection methodology is tested across different energy ranges, and that any phase-resolved period evolution is anchored both in the hard and soft X-ray domains. 

In the later part of the study, we perform a detailed analysis of the non-stationary QPP events presented in the catalogue of Mehta et al. (2023); this choice is motivated by the intent to build upon a well-characterized set of events where non-stationarity has already been clearly detected, rather than performing a broader survey of data from a relatively recent instrument like HEL1OS. 
In their work, Mehta et al. (2023) \cite{Mehta2023} analysed GOES/XRS soft X-ray lightcurves for 98 M- and X-class flares in Solar Cycle 24 using FFT-based method and identified non-stationary QPPs with measurable period evolution between flare phases.
The goal of our work is to use that to extract a richer set of quantitative parameters on QPP characteristics, using a method described in Section \ref{subsect:methods}, which targets non-stationary features, to facilitate a deeper physical interpretation. By applying the method to this existing catalogue, we extract several key properties for each QPP event: (1) its timing and location relative to the flare's impulsive and peak phases, (2) the duration of the QPP signal, (3) the presence of multiple modes, and (4) the magnitude and direction of the period drift.

\subsection{Methods}
\label{subsect:methods}

The first objective of this study is to establish a rigorous approach for detecting and characterizing non-stationary quasi-periodic pulsations in solar flare lightcurves, focusing first on some of the recent observations from CdTe and CZT hard X-ray channels of HEL1OS. The second objective is to demonstrate the utility of the HEL1OS instrument in studying QPPs in the X-ray regime. The strategy we adopt combines VMD with HHT analysis and the wavelet transform, thus combining data-driven adaptive decomposition with well-known time–frequency diagnostics. This combination is selected for two complementary needs : 1) a signal decomposition that can faithfully isolate instantaneously narrowband, drifting oscillatory components in noisy, flare lightcurve, and 2) a statistical framework that can demonstrate significance against red-noise backgrounds and permit comparison with previous large-sample QPP studies.

The starting point for our methodology is the recognition, recently reinforced by Mehta et al. (2023), that non-stationarity is the dominant characteristic of flare QPPs. In their systematic analysis using GOES data, more than eighty percent of QPP detections displayed measurable period drift over the course of the flare. Both increasing and decreasing drifts were observed, often confined to one phase of the flare or extending across impulsive and decay intervals. Importantly, the authors noted that such behaviour renders Fourier-based approaches inadequate, since these methods implicitly assume strict stationarity. They pointed out that by repeating this analysis with some method that has time resolution, such as a continuous wavelet transform (CWT) or EMD we may be able to uncover valuable information about the time evolution of the apparent period drifts. And that it may also be useful in discerning the generation mechanism(s) that are active in the appearance of these QPPs. Our approach is therefore designed as a natural progression of that work. We use their catalogue as ground truth, and apply the combined VMD–HHT–CWT methodology to the same flares in order to estimate when and how the detected QPP periods drifted, and whether multiple oscillatory components, if present, can be separated and tracked across flare phases. 
\vskip6pt

\textbf{Variational mode decomposition}
\vskip6pt
The decomposition stage of our analysis relies on VMD, which was developed as a mathematically well-posed alternative to EMD. While EMD and its ensemble variants have been widely applied in solar flare QPP studies \cite{Kolotkov2015,Nakariakov2018}, their reliance on iterative sifting can lead to mode mixing and sensitivity to noise. Consequently, physically distinct oscillations may split across multiple modes or merge with unrelated fluctuations. This can adversely affect the results in flare QPP analysis, in flares where strongly anharmonic waveforms are common. VMD addresses these deficiencies by recasting the decomposition as a constrained variational problem. The algorithm concurrently seeks a predefined number of modes, each of which must be spectrally compact and collectively reconstruct the signal. The optimization is solved in the Fourier domain using the alternating direction method of multipliers. This formulation provides two decisive advantages. First, by enforcing spectral compactness, VMD prevents contamination of one mode by wide-ranging timescales, thereby mitigating the mode mixing problem. Second, the Fourier-domain implementation is mathematically equivalent to Wiener filtering, which renders VMD substantially more robust to noise. In practice, this means that the modes extracted by VMD are narrowband, stable, and well separated, making them suitable for subsequent Hilbert spectral analysis. Once the signal is decomposed by VMD, we apply the Hilbert transform to each extracted mode, yielding analytic signals from which instantaneous frequency and amplitude are obtained. This produces a Hilbert spectrum, effectively a time–frequency representation derived from the data-adaptive decomposition. The crucial condition for the validity of this step is that each mode represents a genuine mono-component oscillation, a condition that VMD approximates well but EMD often violates. The Hilbert spectrum is therefore an appropriate tool for quantifying the time evolution of QPP periods and amplitudes, particularly in the impulsive and decay phases where drift and damping are expected.

Complementing the VMD–HHT analysis, the wavelet transform is applied to a segment of the lightcurve that displays enhanced power in the Hilbert spectrum for a given intrinsic mode. Applying the wavelet transform to the entire lightcurve can drown out the sometimes feeble periodicities in the low-frequency oscillations, and thus a segmented lightcurve is used. The standard continuous wavelet transform provides a familiar and widely accepted view of the time–frequency distribution of power, with analytic red-noise significance testing available following Torrence and Compo (1998). In this work, the wavelet analysis is used not only as a confirmatory check of the modes extracted by VMD, but also as an independent way of visualizing non-stationarity and benchmarking the Hilbert spectrum.

Mehta et al. (2023) highlighted specific limitations in current methods and hinted at possible improvements : first, that Fourier and periodogram-based analyses cannot accommodate drift in period; second, that EMD-derived instantaneous frequencies are compromised by mode mixing and noise; and third, that wavelet significance tests are conservative but may still misclassify transient or drifting signals. Considering all these factors, to address each of these shortcomings, we implemented a QPP search method by integrating VMD decomposition, which explicitly suppresses mode mixing, with Hilbert spectral analysis, which yields well-defined instantaneous frequencies, and by complementing both with wavelet analysis.The workflow was implemented using custom analysis scripts built around the standard algorithms available within the Signal Processing Toolbox of Matlab R2024a.

Though VMD has several advantages, it is not without limitations. The decomposition depends on user-defined hyperparameters, particularly the number of modes and the bandwidth penalty parameter $\alpha$. Overestimating the number of modes can lead to artificial splitting of a single physical oscillation, while underestimating the number of processes may merge distinct processes into one component. Similarly, the choice of $\alpha$ controls the spectral compactness and can influence the stability of the extracted frequencies. Although we employed physically motivated parameter selection and consistency checks with wavelet scalograms, systematic hyperparameter optimisation remains a challenge for large-sample applications. 

\begin{figure}[h!tbp]
        \includegraphics[width=1.0\textwidth]{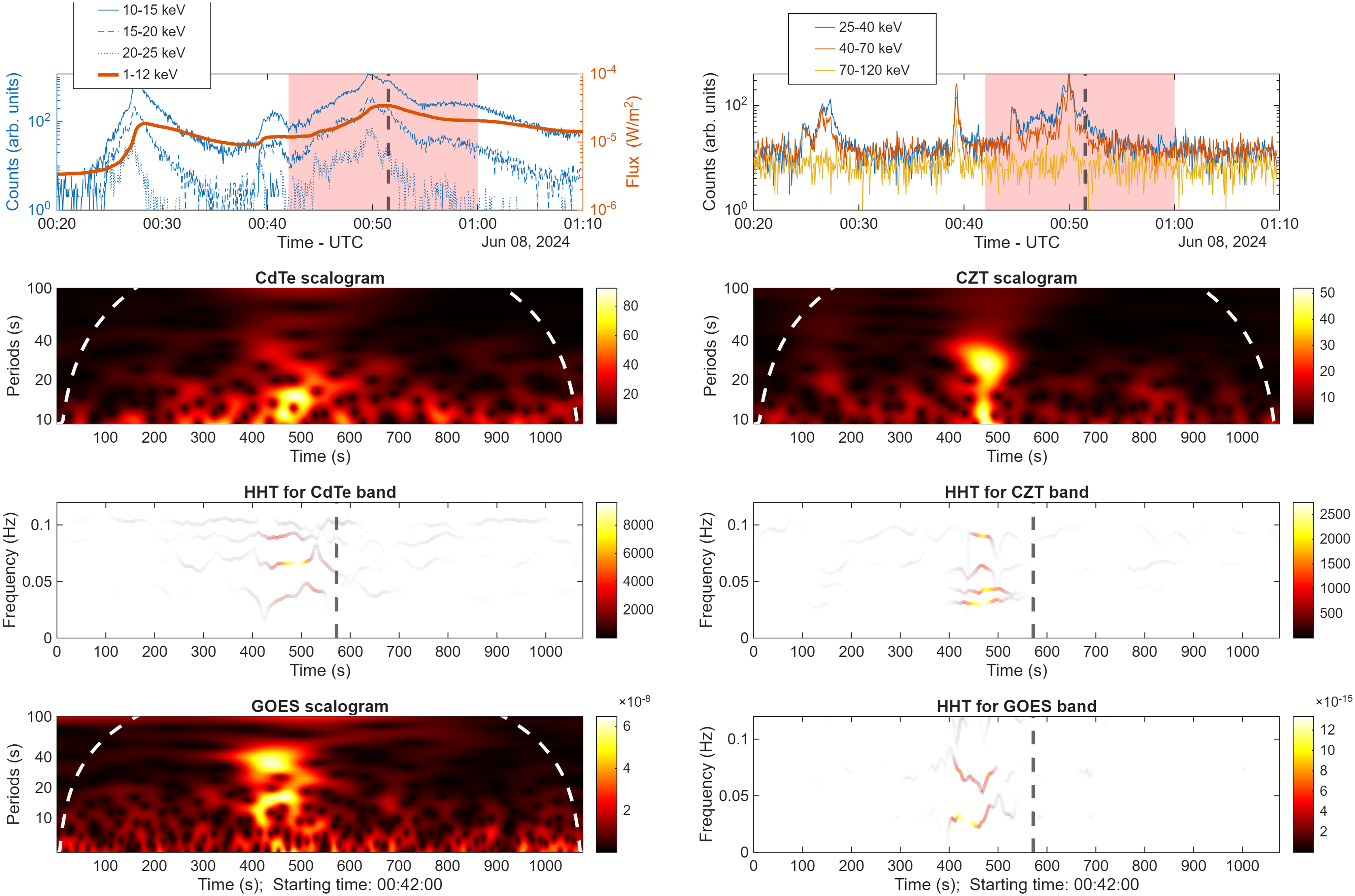}
	\caption{Shows the analysis of the M3.3 solar flare on 8 June 2024 observed with HEL1OS and GOES. Panels in the top row show multi-band hard X-ray lightcurves from the HEL1OS CdTe and CZT detectors together with the GOES soft X-ray flux, with the flare extent used shown as the region shaded in pink. The vertical dashed black line indicates the flare peak in  GOES soft band. The second row displays the wavelet scalograms computed from the integrated CdTe (10–25 keV) and CZT (25–120 keV) lightcurves, with the cone of influence indicated by the dashed white curves. The third row's panels show the HHT spectra derived from variational mode decomposition, revealing time-localized oscillatory ridges. The last row displays the GOES soft band scalogram and HHT spectra for the same duration. The analysis identifies transient QPP modes with periods of approximately 12–20 s and 30–40 s occurring primarily during the impulsive phase of the flare.}
	\label{fig:Event analysis 1}
\end{figure}

\section{Data analysis and Results}
\label{sect:analyss_and_results}

\subsection{ HEL1OS case studies}
\label{subsect:case_studies}

We begin by presenting three representative flares observed with HEL1OS, analysed in parallel with one-second cadence GOES soft X-ray data. These examples illustrate the range of non-stationary and multi-modal behaviours that our methodology is designed to capture, and they can serve as templates for the classification scheme applied later to a larger sample. For each flare, lightcurves from the HEL1OS CdTe and CZT detectors were decomposed using variational mode decomposition, Hilbert spectra were constructed from the extracted modes, and results were cross-checked with wavelet transform of the segmented lightcurve. To ensure that edge effects do not dominate the output and the detected periods reside within the cone-of-influence, the segmented lightcurve is extended appropriately on either side.  GOES lightcurves were analysed with the same procedure to provide an independent thermal reference in the soft X-ray band.


The solar flare of 2024-June-8th, spanning the time 00:42 to 00:57 UT, GOES class M3.3 was analysed using data from the HEL1OS CdTe and CZT detectors, alongside soft X-ray data from the GOES XRS instrument. Figure \ref{fig:Event analysis 1} shows the HEL1OS lightcurves across multiple non-overlapping energy bands with a cadence of 4 seconds, while GOES lightcurve from the long wavelength band (1-8 \AA), was processed using its time derivative to enhance impulsive features.  Variational mode decomposition was applied using six modes to isolate intrinsic mode functions (IMFs), excluding the first, which is dominated by noise, and the sixth, which represents a slow-varying trend. The remaining IMFs (2–5) were subjected to Hilbert spectral analysis to extract instantaneous frequency and amplitude trends as the flare progresses. Periodic ridge segments were identified where the amplitude exceeded the 95th percentile and the duration spanned at least three full cycles of the local period, ensuring robust QPP detection. Complementary wavelet (CWT) analysis was performed using the Morlet mother wavelet. The statistical significance of peaks in the global wavelet spectrum was evaluated assuming a red-noise background described by a first-order autoregressive AR(1) process. Periods exceeding the 95\% confidence level were considered significant.

The analysis revealed a set of QPP modes predominantly during the impulsive phase of the flare, from approximately 400 to 550 seconds from the considered start time. In both GOES and CZT detectors, a long-period mode with periods around 30-40 seconds was observed across nearly all energy bands, with onset times progressively delayed at higher energies, suggesting an energy-dependent modulation. Short-period modes (12-20 seconds) also appeared transiently across energy bands, with the highest visibility in lower energy channels GOES and CdTe. Notably, as can be observed from the HHT spectra and scalograms, the QPP periods obtained from GOES have a greater overlap with the CZT bands than CdTe. The timing or the temporal location of these QPPs, in all the energy bands, occurs in the impulsive phase of the flare before the peak of the GOES flux. Although there are distinct oscillatory modes detected in the impulsive phase, there is no significant presence of the same modes or generation of newer modes in the decay phase of the flare. The presence of multiple coexisting periodicities in the hard and soft X-ray bands, especially with their co-temporal nature, supports the interpretation that magnetohydrodynamic (MHD) wave trains or modulated particle acceleration processes are responsible. 

\begin{figure}
	\centering
        \includegraphics[width=1.0\textwidth]{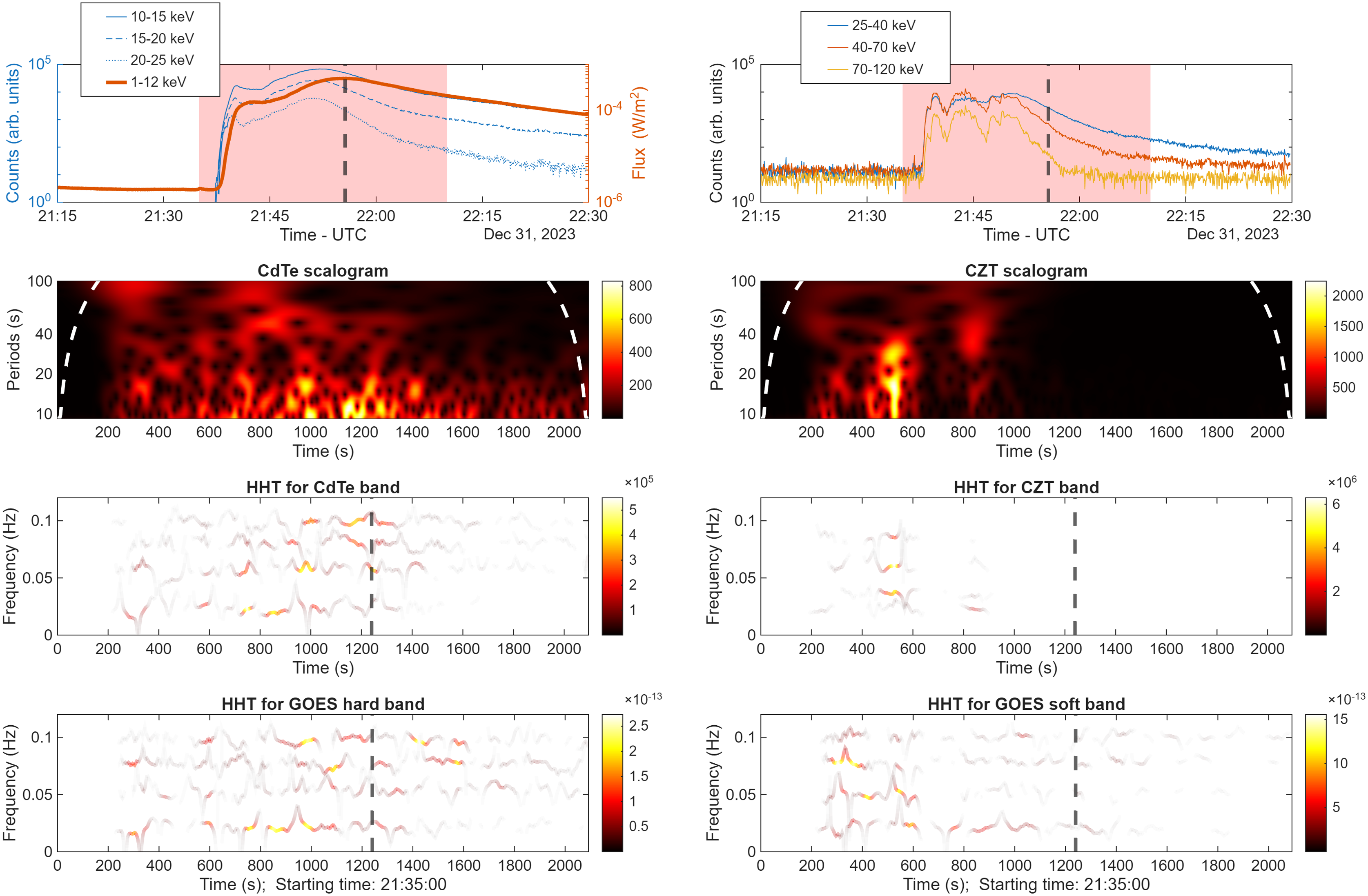}
	\caption{Shows the analysis of the X5 flare on 31 December 2023.
		HEL1OS CdTe and CZT lightcurves and GOES soft X-ray flux are shown in the top row's panels in the same format as Figure 1.  Wavelet scalograms of CdTe and CZT energy bands in the second row reveal transient oscillatory power in the 10–18 s period range. In the third and fourth rows HHT spectra of integrated CdTe, CZT, GOES hard, and GOES soft bands are plotted to indicate similar oscillatory modes' episodes occurring intermittently across energy bands.}
	\label{fig:Event analysis 2}
\end{figure}


The second flare, on 2023-December-31st, exhibits a different pattern. This is an X5 limb flare that started at 21:37 UT.  Quasi-periodic power is first observed in the GOES and CZT lightcurves during the impulsive phase, with well-defined ridges in the Hilbert spectra in the range of 10-18 seconds. In contrast, the CdTe data only begin to show bursts of oscillations ten minutes later, extending into the early decay phase, as shown in Figure \ref{fig:Event analysis 2}. Both the wavelet scalograms and the HHT spectra show the temporal separation of QPP modes across bands, where the GOES and CZT bands show prominent modes at 400 - 500 seconds from start time, while the modes detected in the CdTe energy channels only show up at 1000 seconds. The oscillatory signatures appear to be fragmented into short-lived ridge segments rather than forming a continuous track across the flare peak. This suggests that while the characteristic timescale may be comparable across emission bands, its manifestation is intermittent in space and/or time. The similar period range combined with fragmented timing implies a quasi-periodic driver whose radiative signature is selectively expressed in different plasma components.

\begin{figure}
	        \includegraphics[width=1.0\textwidth]{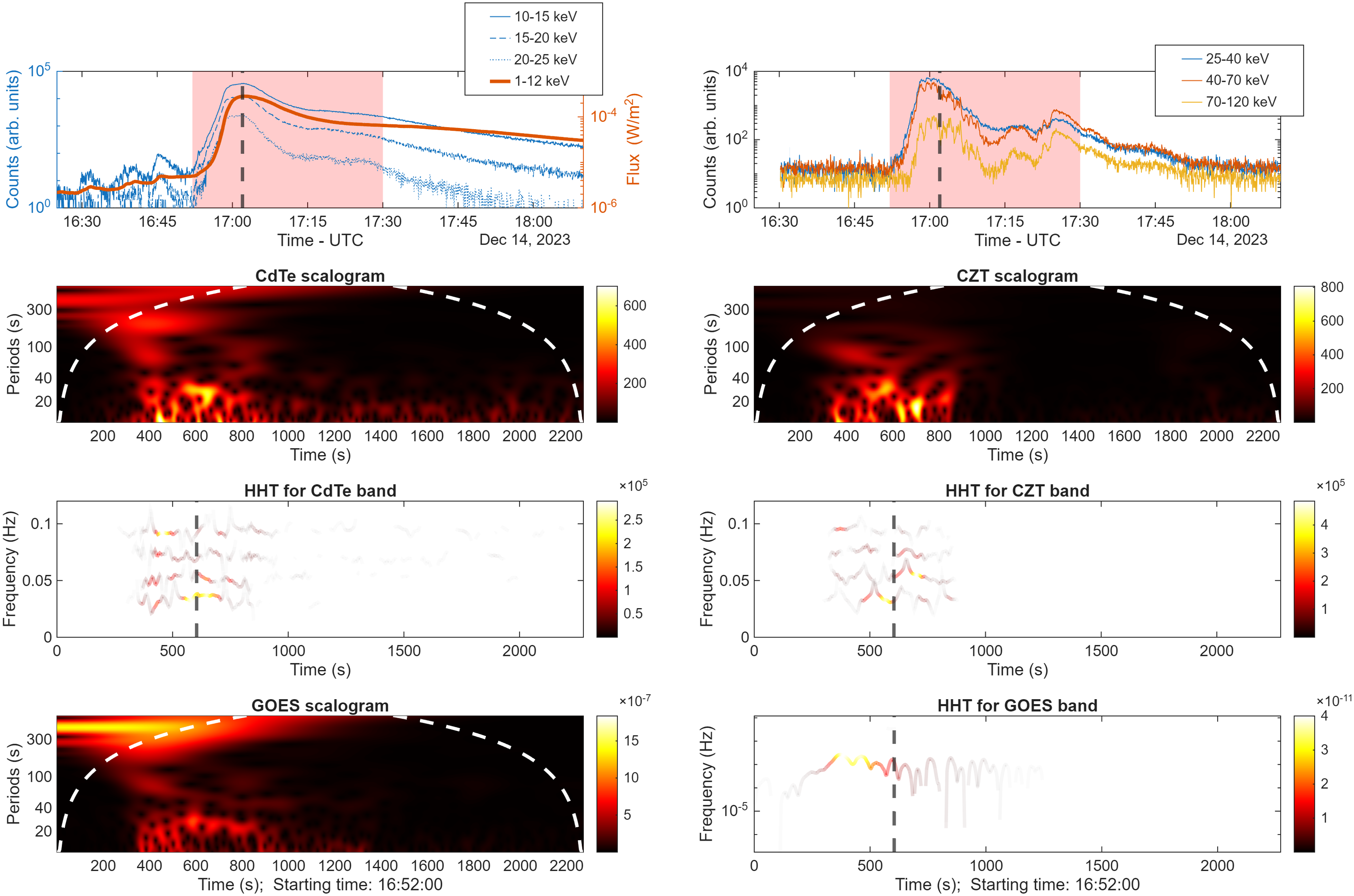}
\caption{Shows the analysis of the X2.8 flare on 14 December 2023 illustrating pre-flare oscillatory behaviour. The panel format is similar to that of Figure 1 -  HEL1OS lightcurves and GOES soft X-ray emission are shown together with wavelet scalograms and HHT spectra. HHT spectrum for the GOES band is plotted with a log-scaled frequency axis to demonstrate a dominant long-period oscillation ($\sim 6.5$ min) that appears prior to the impulsive phase and persisting into the flare onset.}
	\label{fig:Event analysis 3}
\end{figure}

The third flare, which is an X2.8 flare that occurred at 16:47 UT on 2023-December-14th, was studied to explore the impact of preflare activity during the subsequent phases of the flare. The time-window selected for the analysis of the event is between 16:53 and 17:30 UTC.  In this event, the CZT and CdTe bands show a non-stationary mode in the period range of 20-30 seconds. Also observed are shorter periods in the range of 10-20 seconds on either side of the GOES peak. Though hard X-ray lightcurves show statistically significant oscillations demonstrating the presence of multi-modal QPPs, the GOES shows a milder signal trend especially in the wavelet scalogram, shown in Figure \ref{fig:Event analysis 3}. The analysis identifies a dominant periodicity centered between 360 and 400 seconds (approximately 6.5 minutes) that is particularly pronounced in the SXR emission. It is interesting to note that the selected start time (16:53) is later to the NOAA start time (16:47) of the flare, indicating that the impulsive phase is in progress, during which a clear low-frequency oscillatory signature that is remnant of pre-flare oscillations is visible in the low-energy scalograms. A closer look at the GOES and CdTe lightcurves shows three distinct triangular shaped pulses in the pre-flare phase of the flare. The presence of preflare SXR oscillations suggests that coronal plasma was already undergoing quasi-periodic modulation, possibly due to loop oscillations or small-scale reconnection, and that these processes may have acted as precursors or triggers for the impulsive energy release. From the observations, withstanding the edge-effects, it appears that these preflare oscillations may have survived up till and beyond the peak of the flare. The persistence of this spectral feature throughout the impulsive flare phase suggests that the driver of this periodicity is a robust macroscopic structure that remains coherent despite the explosive dynamics of the eruption.

All three flares have QPP detected during or before the impulsive peak of the SXR flux, with no prominent, independent modes in the decay phase especially away from the flare peak. Could this observation be due to the methodology used, or is it due to the absence of thermal QPPs, or are thermal QPPs in the decay phase of flares extensions of their impulsive phase counterparts? We aim to explore these questions by applying the above method for a statistical study on the existing non-stationarity catalogue to primarily extract the location and extent of the QPPs. Altogether, the HEL1OS case studies demonstrate that non-stationary QPPs manifest in several distinct temporal relationships between thermal and non-thermal flare signatures. These categories may form a basis for a larger statistical classification of QPPs based on their properties of non-stationarity.

\subsection{Analysis of the flares in non-stationary QPP catalogue}
\label{subsect:catalog_results}

The framework was then applied to the catalogue of 98 M- and X-class flares analysed by Mehta et al. (2023). First, for each event, we identified the flare phase during which the QPP occurred, determined whether one or multiple oscillatory modes were present, and estimated period drift across phases of the flare. Where multiple modes appeared, we distinguished between genuinely multi-modal behaviour (simultaneous modes with distinct periods) and sequential modes (independent oscillations appearing at different times). From the case studies discussed above it is apparent, in all three flares, that no significant oscillatory mode is detected in the far decay phase and that any detected mode has been an extension from the impulsive phase. Similar results are observed even in this case, when the entire lightcurve is used for the analysis, mainly attributable to the contrast in modulation factors during impulsive and decay phase oscillations \cite{Hayes2016}. When the entire lightcurve is considered, the post-peak phase after the end of non-thermal emission shows a much lower power due to the switch to thermal regime. This essentially makes it difficult to detect decay phase oscillations when using the entire lightcurve. Thus, the decision to bifurcate the flare lightcurve into impulsive and decay halves by Mehta et al., though based on a matter of convenience, is a pertinent one. But, as the non-thermal emission need not end at the soft X-ray peak \cite{Veronig2002}, we split flare lightcurves into two with the impulsive part slightly extending into the decay part beyond the flare peak. The same method is applied to both these lightcurves to find QPPs. For each flare, we measured the period(s), relative timing, and duration  of significant modes, and the flare phase classification. After requiring statistically significant detections in both the impulsive and decay segments, 74 flares were retained for detailed two-phase comparison. The normalized flare phase was defined as \begin{equation} \phi = \frac{t - t_{\mathrm{start}}}{t_{\mathrm{end}} - t_{\mathrm{start}}}, \end{equation} with $\phi = 0$ at the start of the flare, $\phi = 0.5$ at the GOES peak, and $\phi = 1$ at flare termination. For each detected mode, we measured the central phase $\phi_c$, its duration, and the corresponding period in the impulsive ($P_i$) and decay ($P_d$) segments. The evolution of the period across the peak was quantified as \begin{equation} \Delta P = P_d - P_i. \end{equation} To investigate how QPP periods evolve during flare development, we analysed the events in a phase–period–duration parameter space. From the set of impulsive- and decay-phase oscillations identified in the analysis for each flare, only those modes that closely match the values from the catalogue were considered, resulting in 74 events. Of the 74 flares with two-phase detections, 42 events (56.8\%) exhibit small absolute drifts $|\Delta P| < 10$~s, while 32 events (43.2\%) show larger period differences.  

\begin{figure}
	\centering
    	        \includegraphics[width=1.0\textwidth]{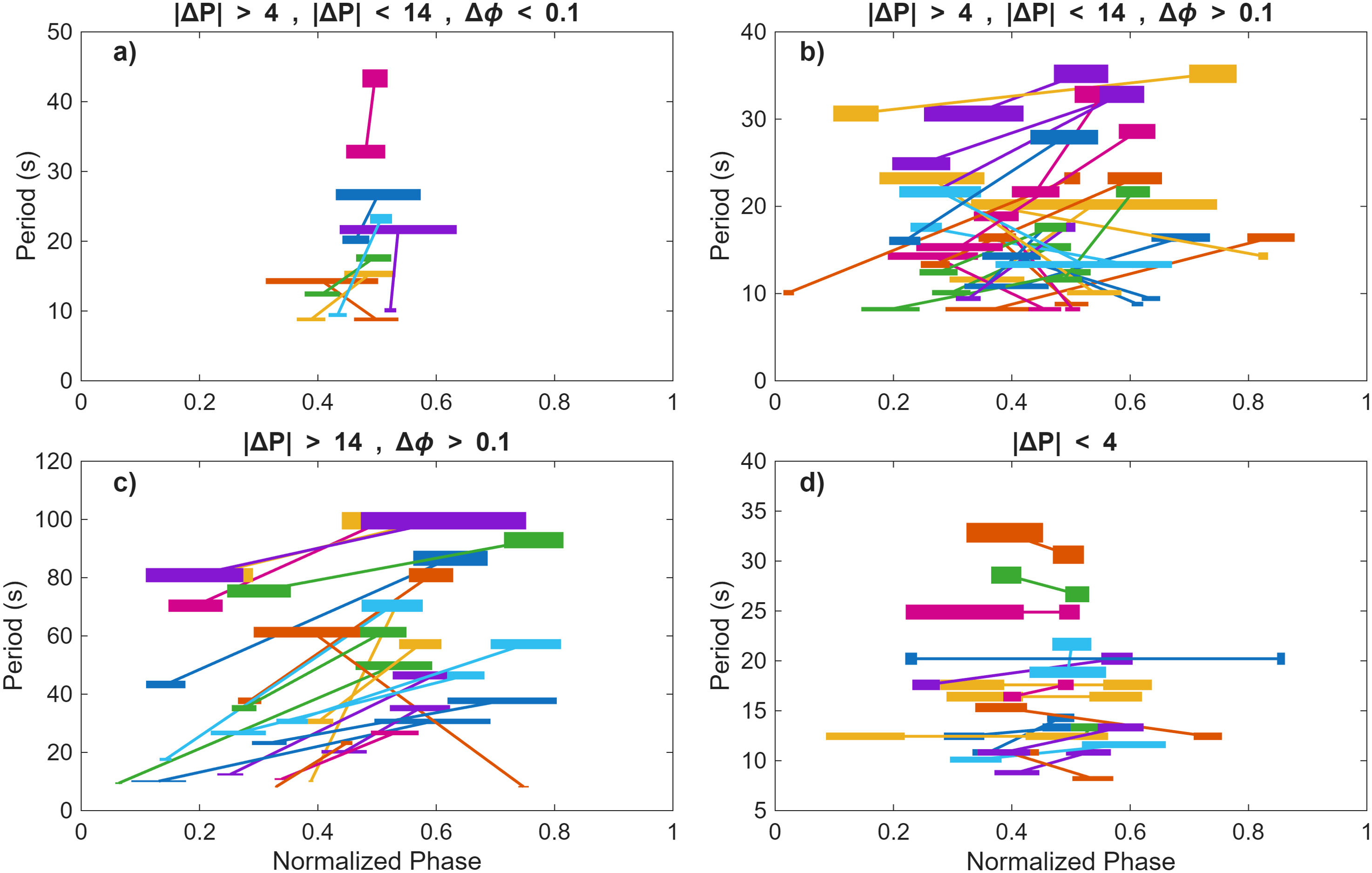}
	\caption{Shows the phase–period representation of QPP evolution across the flare sample.Each panel shows the central phase and period of oscillatory episodes detected in the impulsive and decay phases, with horizontal bars indicating the temporal width of each QPP interval. Lines connect the two phases for each flare. Four regimes are identified:
		(a) overlapping multi-mode oscillations with moderate period drift,
		(b) intermittent oscillations with similar periods but separated in phase,
		(c) strongly non-stationary events exhibiting large period changes, and
		(d) nearly stationary oscillations with negligible drift.}
	\label{fig:phase_period_dist}
\end{figure}

Events can be separated into four populations based on the values of the period drift $|\Delta P|$ and the phase separation $\Delta\phi$ between oscillatory episodes, as shown in Figure \ref{fig:phase_period_dist}. The phase centroids of each QPP are connected in the phase–period plane, with the horizontal extent representing the duration of the oscillatory episode. A small group of events (7 flares) exhibit moderate period drift ($4 < |\Delta P| < 14$~s) but very small phase separation ($\Delta\phi < 0.1$). In these cases, the oscillatory intervals overlap significantly in time across the impulsive and decay phases. The periods remain comparable, but multiple ridge segments coexist within the same phase window, suggesting the simultaneous presence of more than one oscillatory component. The largest population (29 flares) shows moderate period drift ($4 < |\Delta P| < 14$~s) but significant phase separation ($\Delta\phi > 0.1$). In these events, the oscillatory episodes occur at similar periods but are separated in time. The ridges therefore do not overlap across flare phases even though the period scale remains comparable. A third population (20 flares) displays a large period drift ($|\Delta P| > 14$~s) together with significant phase separation ($\Delta\phi > 0.1$). These events exhibit pronounced changes in the dominant period between oscillatory episodes, with periods spanning a wider range that extends to $\sim100$~s in some cases. Finally, 18 events exhibit very small drift ($|\Delta P| < 4$~s), representing nearly stationary oscillations. In these flares, the dominant period remains approximately constant over the observed oscillatory interval. Across all populations, the central locations of the oscillatory episodes cluster preferentially around $\phi\approx0.3$–$0.6$, indicating that QPP activity is detected most frequently around the impulsive–peak transition of the flare. However, the oscillatory episodes themselves may extend across a wider phase range, particularly in the intermittent and multi-modal populations.

\begin{figure}
	\centering
    \includegraphics[width=1.0\textwidth]{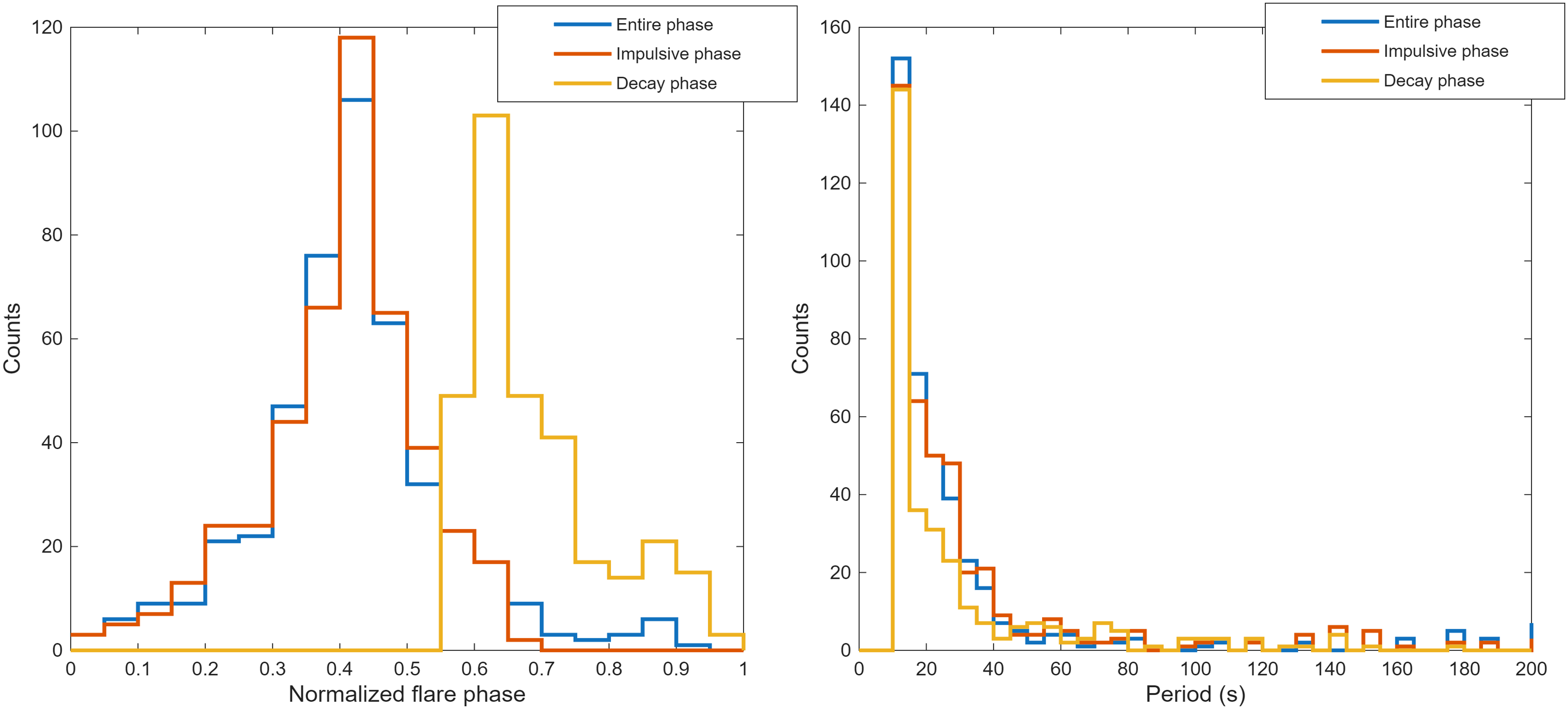}
	\caption{Left: Shows the distribution of QPP central locations with respect to normalized flare phase.Histograms shown are for all the significant modes QPPs detected using the full flare lightcurve, the impulsive phase only, and the decay phase only. The distributions demonstrate that most QPP activity occurs near the flare peak. The decay-phase analysis shows a relative enhancement of longer-period oscillations at later phases. Right: Shows the distribution of QPP periods for all the significant modes detected for the three different lightcurve-extents used in the analysis.}
	\label{fig:All phase hist}
\end{figure}

The $\Delta P$ distribution further reveals an asymmetry: positive drifts are substantially more common than negative ones. Positive drifts ($\Delta P > 0$) dominate, occurring in 73.0\% of events, whereas negative drifts account for 20.3\%.  Large negative drifts are rare and typically of smaller magnitude than positive drift cases. This systematic tendency toward an increasing period across the flare peak suggests that characteristic timescales generally lengthen as the flare evolves. Considering all the significant modes detected for each phase of flare lightcurves, histograms were made to find the most probable phase location for the oscillations. Both the impulsive-only and full-duration QPP histograms produce nearly identical distributions strongly concentrated around $\phi \approx 0.45$, indicating that the same oscillatory population is detected in both treatments, as shown in Figure \ref{fig:All phase hist}. The decay-only lightcurve analysis, however, shows a relative enhancement of longer-period oscillations at $\phi > 0.6$, consistent with the emergence of a secondary population in the late decay phase\cite{Hayes2020}.


\section{Discussion}
\label{sect:disc}

Three flares (2024‑Jun‑8, 2023‑Dec‑31, and 2023‑Dec‑14), which are observed by the HEL1OS CdTe and CZT detectors in multiple energy bands, were analysed to understand QPP evolution bandwise and within a band. Each lightcurve was decomposed using a six‑mode variational mode decomposition; the first high‑frequency mode (noise) and the sixth long‑term trend were discarded . The HHT analysis provided instantaneous periods for the remaining IMFs, and CWT with significance tests confirmed the non‑stationary nature of the oscillations. The detected ridges were retained only when their amplitude exceeded the 95th percentile, and they persisted for at least three cycles, ensuring that only genuine QPPs were reported.

The three flares analysed here can be empirically segregated into three different categories, based on the timing of the detected periodicities. The first event comprises those in which QPP ridges are co-temporal in the HXR and SXR bands, with similar periods persisting across both energy ranges. In such cases the oscillatory modulation appears to act simultaneously on the non-thermal acceleration signatures and the thermal response, consistent with a global modulating mechanism such as standing magnetohydrodynamic oscillations or oscillatory reconnection that couples directly to both channels. 

Event 2 provides direct observational evidence for intermittency. Although the dominant period range remains comparable across bands, the ridge timing is fragmented and energy-dependent. The CdTe modulation aligns preferentially with GOES hard band emission, while the CZT modulation correlates more closely with GOES soft band emission. The flare likely undergoes spatially localized or temporally bursty quasi-periodic acceleration episodes. This behaviour is consistent with bursty or oscillatory reconnection in fragmented current sheets, where plasmoid formation or tearing-mode dynamics generate quasi-periodic but spatially localized acceleration episodes \cite{Kliem2000,Nakariakov2010,McLaughlin2018}. In these scenarios, the characteristic timescale may be governed by local Alfv\'enic or tearing-mode dynamics, while the observable emission depends on evolving plasma conditions and energy partitioning. This demonstrates that the characteristic timescale is decoupled from continuous emission, supporting a scenario in which the underlying driver is persistent but observationally intermittent.

\begin{figure}
            \includegraphics[width=1.0\textwidth]{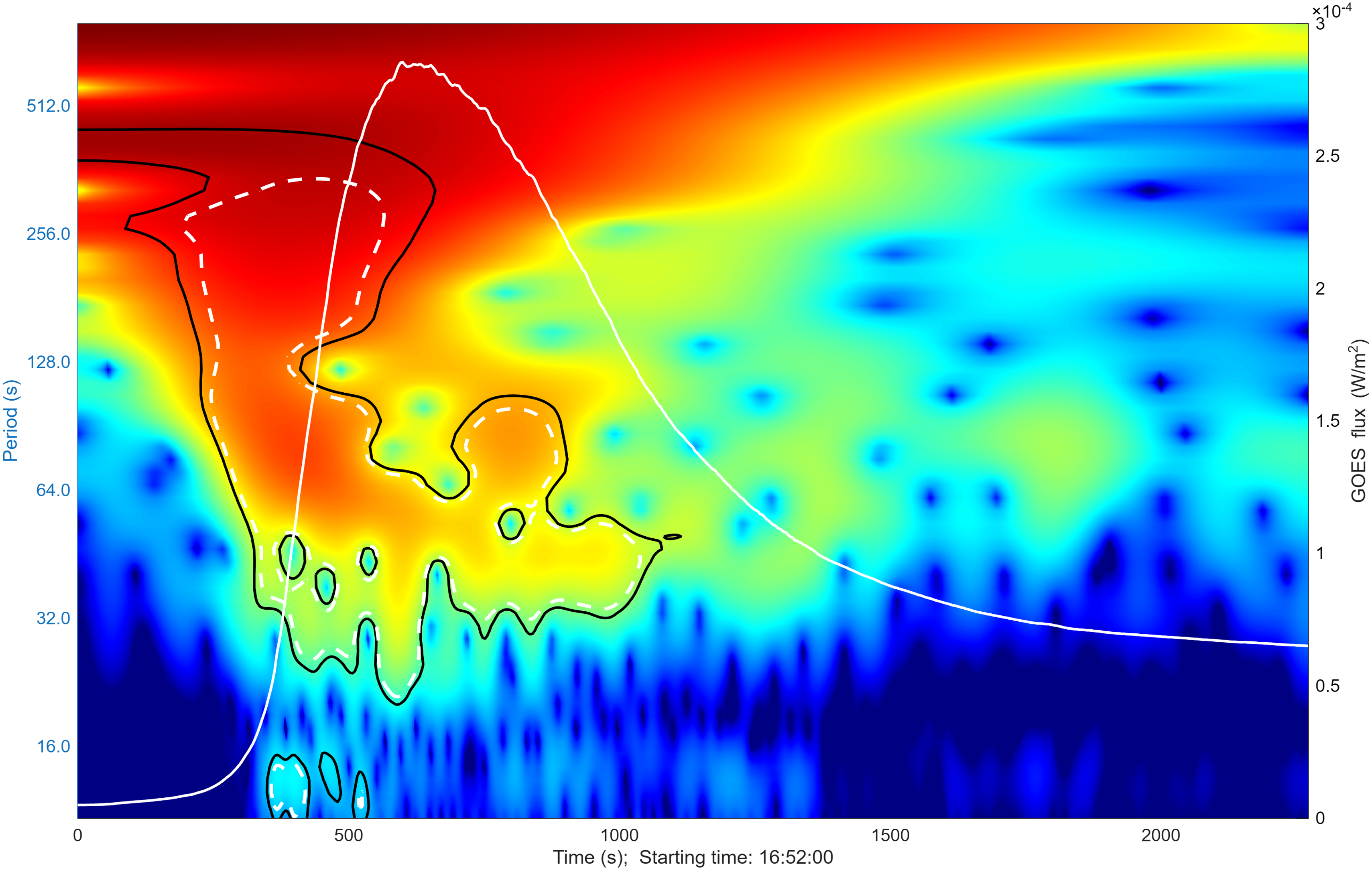}
	\caption{Shows the wavelet scalogram of the sixth IMF of the GOES lightcurve of the flare on 2023-December-14th, illustrating the evolution of a pre-flare oscillation into the impulsive phase. The scalogram shows a long-period mode ($\sim6.5$ min) present prior to flare onset that persists into the early impulsive phase. The solid black and dashed white contours mark the 95 and 99\% significance levels respectively.The grey line plot shows the GOES lightcurve for the same duration of the scalogram. }
	\label{fig:Proof plot}
\end{figure}

The third event is characterized by pre-flare SXR oscillations that precede the onset of HXR QPP ridges. These preflare modulations are not detected in the soft X-ray background as statistically significant ridges in the Hilbert and wavelet spectra, due to the constraints imposed in the procedure, but are captured well in the smoothened lightcurve. These are followed by the appearance of hard X-ray oscillations once the flare's impulsive acceleration phase begins. In some cases the SXR pre-flare oscillation may continue into the impulsive phase, in others it is replaced by distinct HXR oscillations. The former is the case for the December 14th, 2023 flare. In a recent study by Zimovets et al. (2025) \cite{Zimovets2025} clearly identified a $\sim6.5$-minute pulsation in the pre-flare emission of the same event, interpreting it as a manifestation of repetitive magnetic reconnection as a loading-unloading cyclic process of supply and transformation of free magnetic energy into kinetic energy of plasma particles in the current sheet. They also established that pre-flare pulsations were emitted from different locations of a complex magnetic structure of the active region, but not from a single loop. The noteworthiness of the present analysis is the observational confirmation that these pre-flare pulsations survive the catastrophic trigger of the flare eruption,which itself oscillates at a higher frequency. The pulsations appear to be highly non-stationary as the period rapidly evolves from its pre-flare value of $\sim6$ to $\sim1$ minutes, during the impulsive phase as shown in Figure \ref{fig:Proof plot}. Recent studies by Song et al. (2025) \cite{Song2025} have characterized the rapid sub-minute quasi-periodicities (11 and 20 seconds) driven by local reconnection dynamics in this flare, which also have been corroborated in this work. This implies that the large-scale magnetic resonator -- likely a overarching loop arcade -- retains sufficient topological stability to sustain its oscillation mode, thereby modulating the reconnection rate even as the core magnetic field undergoes violent reconfiguration.  

\subsection{Period Drift and Interpretation}
\label{subsect:period_drift}

By tracking the instantaneous period along each significant ridge in the Hilbert spectrum and confirming it with mode-wise wavelet scalograms, we quantified the evolution of QPP timescales across flare phases. Though we did not find exact QPP periods in both phases for all of the 98 flares in the catalogue, the  trends (in terms of percentage) obtained are quite similar. Of the 74 two-phase events analysed, a large majority of them show small period drifts. Such cases are named weakly non-stationary, where a single oscillatory process is expected to persist while its characteristic timescale evolves gradually. For standing MHD modes in coronal loops, the period scales approximately as $P \sim 2L / nC_{\mathrm{ph}}$, where $L$ is the effective loop length, $n$ is the longitudinal harmonic number, and $C_{\mathrm{ph}}$ is the relevant phase speed  \cite{Edwin1983,Nakariakov2005}. While the simple standing-mode interpretation is often expressed through loop length and phase speed, certain sausage-mode regimes can depend on loop’s minor radius \cite{Nakariakov2012}. Here, we interpret the measured periods conservatively as characteristic timescales rather than as unique mode identifications. During flares, loop arcades are observed to expand, reconnect, and undergo density enhancement due to chromospheric evaporation. Gradual increases in loop length or reductions in Alfv\'en speed resulting from increased plasma density naturally produce modest increases in oscillation period \cite{Kupriyanova2020,Hayes2016}. The predominance of positive drifts in our sample is, therefore, consistent with progressive restructuring of the flaring magnetic arcade. Small apparent drifts may also arise from dispersive wave-train behaviour. Another possibility is the propagation of dispersive fast magnetoacoustic wave trains along coronal waveguides. In such cases the observed instantaneous period evolves because different spectral components travel at different group velocities \cite{Nakariakov2004}. The observed instantaneous period can evolve smoothly even if the underlying structure remains unchanged. 
However, analysis of the events demonstrates that a small absolute drift does not necessarily imply a single continuous oscillatory episode. Although a significant fraction of events cluster near the normalized flare peak ($\phi \approx 0.4$--$0.6$), many small-drift events exhibit moderate or even large phase separation across the flare peak.  

To distinguish weak from strong non-stationarity, the events were separated using thresholds in both period drift and phase separation. The boundary between weak and strong drift was taken as $|\Delta P|= 14$ s. The threshold corresponds to the upper boundary of the dominant cluster of small period drifts in the sample and remains broadly consistent with the $\sim10$ s drift scale reported in the previous work. The phase separation threshold was set at $\Delta\phi$ = 0.1, corresponding to oscillatory episodes that occurred within about ten percent of the flare duration. Such cases are treated as temporally overlapping intervals where simultaneous or coexisting oscillatory modes may be present. As a consequence of these thresholds, no events occupy the quadrant defined by $|\Delta P|>14$~s and $\Delta\phi>0.1$. This absence indicates that large period changes are not observed when oscillatory episodes overlap in time, suggesting that strong period drift generally reflects transitions between distinct oscillatory regimes rather than the smooth evolution of a single mode.

The largest population in the catalogue consists of events with moderate period drift ($4<|\Delta P|<14$~s) but substantial phase separation ($\Delta\phi>0.1$). These events likely represent intermittent oscillatory activity, where the same characteristic timescale is repeatedly excited during the flare evolution. Such intermittency may arise from episodic or bursty energy release in the reconnection region. Oscillatory reconnection and plasmoid-mediated current-sheet fragmentation naturally produce quasi-periodic but intermittent acceleration episodes, in which successive bursts occur at comparable timescales, but are separated in time \cite{Kliem2000,Nakariakov2010,McLaughlin2018}. In this scenario, the period remains relatively stable because it is set by local Alfvénic or tearing-mode dynamics, whereas the observable emission appears in discrete packets as reconnection sites are activated intermittently. Thus, the results suggest that weak non-stationarity encompasses at least two physically distinct regimes: (i) continuous, weakly evolving oscillations spanning the flare peak, and (ii) intermittent same-period oscillations whose timing is fragmented across phases or energy channels.

\begin{figure}
	\centering
    \includegraphics[width=1.0\textwidth]{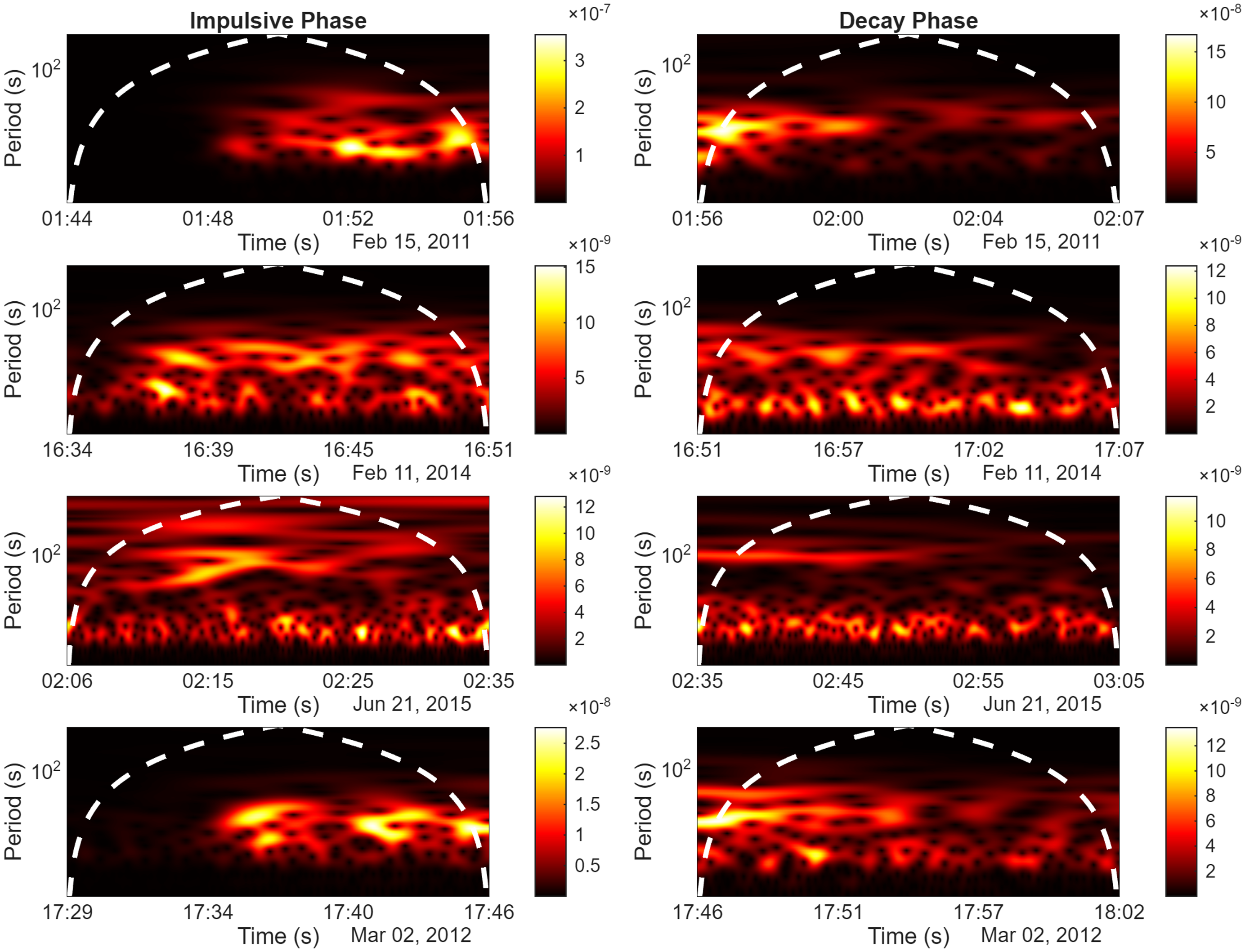}
	\caption{Representative examples illustrating the four regimes of non-stationary QPP evolution identified in the catalogue analysis. For each event, the left and right panels show the wavelet scalograms obtained from the impulsive and decay phases, respectively. The dashed white curves indicate the cone of influence. The upper row shows an event with overlapping oscillatory episodes and small period drift. The second row shows an intermittent event in which similar periods are detected in both phases but are separated in time, suggesting episodic excitation of the same characteristic timescale. The third row illustrates a strongly non-stationary event exhibiting a substantial period change between phases, indicative of a transition between distinct oscillatory regimes or the emergence of a different dominant mode. The fourth row shows an event with weak global period drift between phases. Although the representative periods identified in the impulsive and decay phases remain similar, the wavelet ridges exhibit continuous local evolution within each phase, demonstrating that non-stationarity persists even when the net period change across the flare is small.}
	\label{fig:Four flare categories}
\end{figure}

A smaller subset of events shows overlapping oscillatory episodes with comparable periods. These cases may represent simultaneous oscillatory components within the flare region. Multiple loop systems or wave modes may be excited concurrently, producing overlapping ridge structures with similar period scales. The relatively small fraction of such events suggests that simultaneous multi-mode behaviour is less common than intermittent excitation of similar timescales for events with small drifts in period. To illustrate the physical meaning of the phase–period classification, representative events from each category are shown in Figure \ref{fig:Four flare categories}. While Figure \ref{fig:All phase hist} summarizes the statistical distribution of period drift and phase separation across the catalogue, the wavelet scalograms in Figure \ref{fig:Four flare categories} reveal how these different classes manifest in the time–period domain.

The large-drift subset displays substantial deviations from the unity period ratio, often involving multi-modal QPPs transitions from dominant mode in the impulsive phase to another in the decay phase. In phase–period space, these events appear as separated segments rather than continuous tracks. This behaviour is unlikely to represent smooth evolution of a single standing mode, and instead suggests strong non-stationarity associated with changes in the dominant physical process. One explanation is a transition from reconnection-driven periodic acceleration during the impulsive phase to large-scale MHD oscillations of the post-flare arcade during the decay phase. Oscillatory reconnection models predict that the repetition rate of energy release can evolve as the current sheet geometry and the magnetic stress change over time \cite{Kliem2000,Nakariakov2010}. Load–unload scenarios similarly produce time-dependent recurrence intervals as magnetic energy accumulates and is episodically released \cite{McLaughlin2018}. As the flare progresses and larger magnetic structures become involved, longer-period oscillations of post-flare arcades may emerge \cite{VanDoorsselaere2016,Zimovets2021}. The emergence of longer periods well into the decay phase in the dataset is consistent with such a transition.

Thus, the results indicate that non-stationarity in flare QPPs arises from the interplay of several physical processes: gradual evolution of oscillatory eigenmodes within flaring loop systems, intermittent excitation associated with bursty magnetic reconnection, and transitions between distinct oscillatory regimes as the flare evolves. A phenomenological classification of QPPs behaviour was previously proposed  in the context of microwave observations by Kupriyanova et al. (2010) identified four types of QPPs based on the temporal evolution of their periods, including stationary oscillations, systematic drift toward shorter or longer periods, and multi-period or crossing structures\cite{Kupriyanova2010-ct}. The regimes identified in the present study are consistent with this earlier classification, and extend it by incorporating the phase–period relationship across flare evolution. Particularly, the separation of events based on both period drift and phase proximity allows intermittent and multi-modal behaviour to be distinguished from continuous evolution of a single oscillatory process.

\section{Summary and Conclusions}
\label{sect:conclusion}

This study has examined the non-stationary behaviour of quasi-periodic pulsations in solar flares using a time–resolved analysis framework that combines variational mode decomposition, HHT ridge tracking, and wavelet analysis. By applying the method both to representative HEL1OS flare observations and to a statistical reanalysis of the non-stationary QPP catalogue, we find that the temporal evolution of QPP periods contains important information about the physical processes operating in flaring active regions. The results indicate that QPP non-stationarity is not a single phenomenon but reflects multiple regimes of oscillatory behaviour. Nearly stationary oscillations are consistent with stable eigenmodes of flaring loop systems, while modest period evolution across the flare peak suggests gradual modification of loop parameters as magnetic structures expand and plasma density increases during the flare. A large fraction of events, however, exhibit oscillatory episodes that recur intermittently with similar characteristic periods. This behaviour points toward episodic or bursty reconnection processes in which particle acceleration and energy release occur in discrete bursts rather than as a continuously modulated process. In a smaller subset of events the dominant period changes substantially between flare phases, indicating transitions between distinct oscillatory regimes or the activation of different magnetic structures during flare evolution.This classification extends previous statistical findings, and has scope to be strengthened by identifying the morphological structure underlying period drift using imaging observations of X-rays at a cadence that caters to the observed QPP periods. Methodologically, this study establishes VMD as a powerful tool for decomposing flare lightcurves into spectrally compact narrowband modes that are well suited for Hilbert-based instantaneous frequency analysis. 

From an observational perspective, the results demonstrate the capability of HEL1OS to detect and characterise QPPs across multiple hard X-ray energy bands with sufficient temporal resolution to resolve both impulsive and decay-phase behaviour. Its multi-band sensitivity enables discrimination between non-thermal and thermal signatures of oscillatory processes. In the future, coordinated analysis combining HEL1OS data with SoLEXS'\cite{Sankarasubramanian2025a} soft X-ray observations will enable comprehensive studies of thermal and non-thermal QPP populations over extended intervals of solar activity.

\enlargethispage{20pt}

\ack{We acknowledge the support of U R Rao Satellite Centre and IIT-BHU in carrying out this research work.}


\vskip2pc

\bibliographystyle{RS} 

\bibliography{referen_edited} 
\end{document}